# A Prototype Scalable Readout System for Micro-pattern Gas Detectors [*]

Qi-Bin Zheng (郑其斌)[1,2] Shu-Bin Liu (刘树彬)[1,2] Jing Tian (田静)[1,2]

Cheng Li (李诚) [1,2] Chang-Qing Feng (封常青) [1,2] Qi An (安琪)[1,2]

[1]Department of Modern Physics, University of Science and Technology of China, Hefei 230026, China
[2]State Key Laboratory of Particle Detection and Electronics (IHEP-USTC), Hefei 230026, China

**Abstract:** A scalable readout system (SRS) is designed to provide a general solution for different micro-pattern gas detectors. The system mainly consists of three kinds of modules: the ASIC card, the Adapter card and the Front-End Card (FEC). The ASIC cards, mounted with particular ASIC chips, are designed for receiving detector signals. The Adapter card is in charge of digitizing the output signals from several ASIC cards. The FEC, edged-mounted with the Adapter, has a FPGA-based reconfigurable logic and I/O interfaces, allowing users to choose various ASIC cards and Adapters for different types of detectors. The FEC transfers data through Gigabit Ethernet protocol realized by a TCP processor (SiTCP) IP core in field-programmable gate arrays (FPGA). The readout system can be tailored to specific sizes to adapt to the experiment scales and readout requirements. In this paper, two kinds of multi-channel ASIC chips, VA140 and AGET, are applied to verify the concept of this SRS architecture. Based on this VA140 or AGET SRS, one FEC covers 8 ASIC (VA140) cards handling 512 detector channels, or 4 ASIC (AGET) cards handling 256 detector channels. More FECs can be assembled in chassis to handle thousands of detector channels.

## 1. Introduction

As the fundamental components of high energy physics (HEP) experiments, micro-pattern gas detector (MPGD) offers a great potential as a high resolution particle tracking detector for a variety of applications, especially for the Micromegas [1] and Gas Electron Multipliers (GEMs) [2], which have now reached maturity and become increasingly important. The MPGDs have already been used in large-scale particle physics experiments, like COMPASS [3] [4], ALICE [5] and LHCb [6]. And they are considered as the forward detectors [7-11] thanks to its characteristics such as excellent time and spatial resolution, high radiation resistance, high rate capability, and large active areas.

The widespread uses of the MPGD are a driving factor in the development of corresponding readout electronics. Various applications call for a variety of appropriate readout requirements (signal polarity, radiation tolerance, and a large number of channels, etc.) that urged scientists to establish a scalable multichannel readout system. Therefore, in 2009, the RD51 Collaboration at CERN [12] produced the most potential readout electronics named the Scalable Readout System (SRS) for gas detectors like Micromegas and GEMs, featuring a scalable architecture and a general-purpose chip link interface, allowing the user to choose from a variety of front-end chips on hybrids with integrated spark protection circuitry [13]. Since the first SRS systems were successfully used in ATLAS Micromegas stations in 2010, the readout system progressively gained momentum in the MPGD community. At present, the SRS system has already been considered to be translated into an industrial standard abroad [13], but in China the study on this area is still a blank. This paper presents a prototype SRS system designed by the authors' group.

## 2. Proposed architecture

Referenced the design concept presented by RD51 collaboration, the architecture of this prototype SRS is shown in Fig. 1. The readout electronics mainly consist of three kinds of modules, the analog module called ASIC card which is the nearest part to the detector and measuring charge signals, the analog to digital module named Adapter card sampling the analog module outputs, and the digital module named Front-End Card (FEC) which is responsible for signal processing, data and control commands transmitting-receiving. In a way of "Several-in-One", several particular ASIC cards are connected to one specific designed Adapter through general purpose chip link, and the Adapter is edged-mounted to the fixed FEC through SAMTEC PCIe connectors as a standard 6U*220 mm plug-in. And an AC/DC power module energizes the three kinds of cards. All of the plug-ins and power module are assembled in a 6U chassis. After processing the digitized data, the FEC transmits data to and from server or PC.

[*]Supported by National Natural Science Foundation of China (Grant No. 11222552)
1) E-mail: qbzheng@mail.ustc.edu.cn
2) E-mail: liushb@ustc.edu.cn (corresponding author)

The FEC contains FPGA-based reconfigurable control-logic and I/O interfaces for different specific Adapters. Both the ASIC card and Adapter are working under the control of FEC. So, one FEC can handle a certain number of detector channels. As for the small system, one FEC is sufficient to meet the quantity requirement of detector channels. Furthermore, using Gigabit Ethernet for FEC connections, forming a larger system of more FECs is flexible. A number of FECs would be assembled in parallel to expand the number of detector channels which are readout.

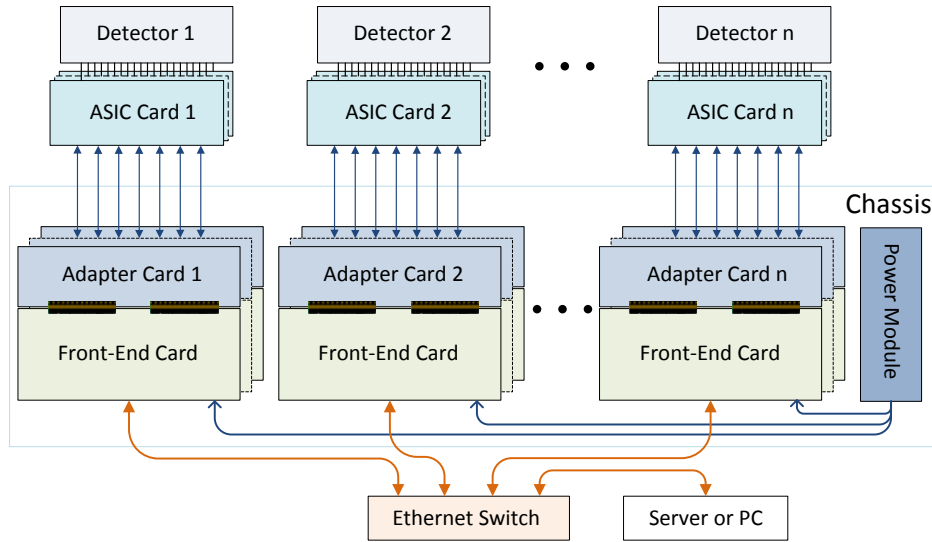

Fig. 1. Block diagram of the proposed SRS.

## 3. Implementation of the SRS

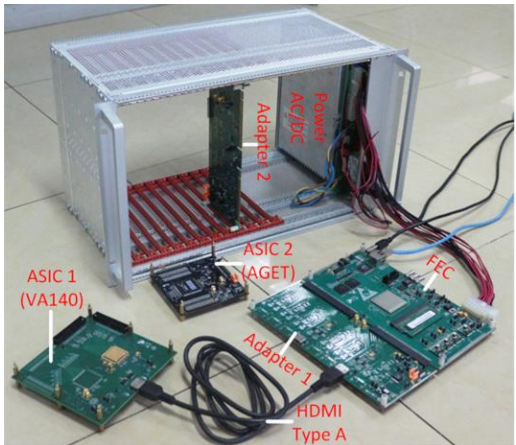

Fig. 2. A photograph of the SRS chassis and its prototype modules.

The prototype SRS system was designed and implemented base on two multi-channel ASIC chips, VA140 [14] and AGET [15] which dynamic ranges were suitable for MPGD. The photograph of the system was showed in Fig. 2. Two sets of ASIC cards with corresponding Adapters were specially designed. The fixed FEC matched the Adapters and formed a standard 6U*220 mm Eurocard which would be assembled in chassis.

### 3.1 The chassis

As shown in Fig. 2, the standard 19-inch chassis which was designed and custom made from SCHROFF with 19 slots at most, was used to assemble the 6U*220 mm plug-ins. Mounted at the rear of the chassis, the stabilized power module (13100-145, SCHROFF) [16] with input voltage ranging from 90 to 264 $V_{AC,}$ supplied 4 high current output powers (+3.3$V_{DC}$ (60 A), +5.0$V_{DC}$ (50 A), +12$V_{DC}$ (12 A), -12$V_{DC}$ (12 A)) for the SRS system.

### 3.2 The FEC card

The FEC module was designed around a Virtex 6 FPGA (XC6VLX240T) as a 6U*120mm card, integrating one DDR3 memory chip, one Ethernet port and general interfaces, as shown in Fig. 3. At the front end, the detector signals were integrated, shaped by the ASIC cards and then amplified, sampled by the Adapter, under the control of the FPGA. The FPGA received the digitized data through SAMTEC PCIe connectors for further processing. In FPGA, data selection algorithms must be developed to meet the requirements of different experiment. The target data was formatted and stored in the DDR3 buffer. For data transmission, we studied the SiTCP [17] which was a hardware-based TCP processor for devices limited by hardware size, such as front-end devices or detectors. The SiTCP processed TCP, IP and Ethernet protocols to realize the process of data transmission. Gigabit-Ethernet media independent interface (GMII) was used to link the Ethernet physical layer device (PHY) and a media access controller (MAC). The detail of the same data transmission based on SiTCP was described in reference [18].

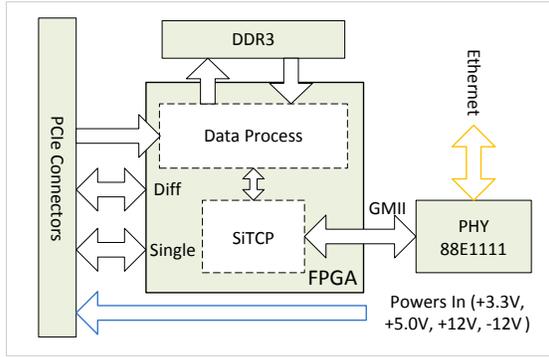

Fig. 3. Block diagram of FEC.

Table. 1. The redundancy design list of PCIe connectors on FEC.

| Types | Numbers | Connected |
|---|---|---|
| Powers | 4 kinds | AC/DC Module |
| Single | 74 wires | FPGA |
| Differential | 34 pairs | FPGA |
| Clock | 1 pair | System Clock |

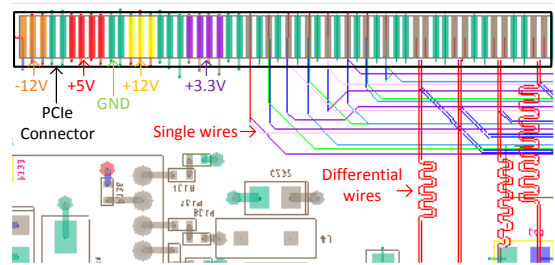

Fig. 4. Part of FEC PCB layout for PCIe connections.

To serve the needs of scalability, the redundancy was fully considered in designing FEC which may cover potential ASIC cards based on different multi-channel chips. Two PCIeX16 connectors were chosen. Both the differential and single IOs of the FPGA were routed to this PCIe interface, with plenty of design redundancy including sufficient IOs, same length of wires (25 mils margin for differential wires, and 50 mils margin for single wires), etc., as shown in table. 1. And Fig. 4 showed part of the PCIe connections, four kinds of power supplies from AC-DC power Module were also connected to the PCIe connector. Through reconfiguring the interface, specific Adapters and corresponding ASIC cards worked normally under the control of the FEC. This architecture expands the system in application dimension. Furthermore, a number of FECs can be assembled in a topology through Gigabit Ethernet port. This method makes the SRS suitable for experiment with large number of detector channels.

### 3.3 Specific ASIC cards and Adapters

The multi-channel ASIC chips such as VA140 [14], AGET [15] and APV25 [19], etc. were applied to verify the concept of this SRS architecture. All of the ASIC cards and Adapters were almost designed in a same structure as shown in Fig. 5 and 6. The input channels of ASIC chips were AC coupled to the detectors and protected against discharge by low capacitance diode arrays (NUP4114UPXV6). Since the ASICs need particular power supplies which were regulated by the local LDOs, the initial voltages were supplied via the HDMI cable from Adapter, as the same way that the ASIC controlling signals and ASIC outputs were transferred between the two cards.

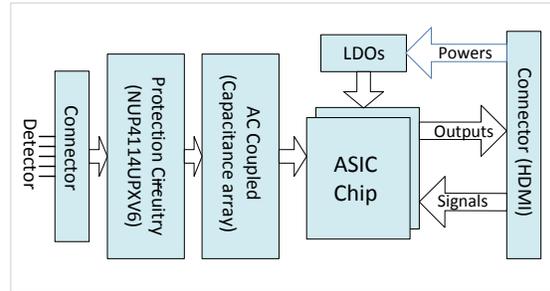

Fig. 5. Block diagram of ASIC card.

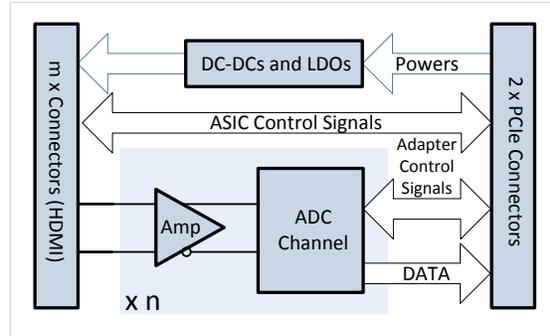

Fig. 6. Block diagram of Adapter card.

The Adapter was a link between the ASIC cards and FEC. As shown in Fig.6, it was designed based on ADCs and tailored to fit the FEC through PCIe connectors. The ASIC and Adapter controlling signals were produced by the FPGA on FEC which also supplies powers to energy them. The particular power supplies needed for Adapters were regulated by local DC-DC converters and LDOs.

### 3.3.1 ASIC cards and Adapter based on VA140

As the design structure and proposal described earlier, the first set of ASIC card and corresponding Adapter was designed based on VA140 chip [14] which is a 64 channels, 6.5 μs peaking time, low-noise, low power, and high dynamic range (from -200 fC to 0 or from 0 to 200 fC) charge sensitive preamplifier-shaper ASIC, and suitable for MPGD and silicon detectors [18]. The overview and partial performance of the chip were described in reference [20].

According to the VA140 datasheet [14], the chip needs 1.5V and -2.0V powers supplies, and 5 single controlling signals (HOLDB, CLKB, SHIFT_IN_B, DRESET, and TEST_ON). The power consumption of VA140 is about 0.2 mW/ch. A 12-bit dual-ADC chips (AD7356) was adopted to digitize the output of VA140, featuring up to 5 MSPS per channel with 4 controlling signals (SCLK, SDATA_A, SDATA_B, and CS) and 1 power supply (2.5V). With the help of some other auxiliary devices, the VA140 ASIC card and Adapter were implemented. It should be pointed out that the VA140 chip in this project was encapsulated in the CQFP132 package. Limited by the packaging technology, only 32 channels per chip were wire bonded out.

Table. 2. Respective requirements for FEC

| Requirements | VA140 | AGET | APV25 |
| --- | --- | --- | --- |
| Powers | 3.3V, 5V | 3.3V, 5V | 3.3V, 5V |
| Single | 72 | 28 | 36 |
| Differential | 0 pairs | 19 pairs | 27 pairs |
| Channels | 512 | 256 | 1024 |

In the VA140 based SRS, the FEC with corresponding Adapter was responsible for digitization of the front-end signals and acted as a carrier for the "8-in-1" card. 8 VA140 ASIC cards, each integrating 2 VA140 chips, read out 512 channels of detector and connected to one Adapter which integrating 8 ADC chips (AD7356). Each channel integrated its eventual signal for 6.5 μs. After the peak was reached (6.5 μs), an external 'HOLDB signal should be applied to sample the value. Immediately after this a sequential read-out can be performed by activating the output bit-register using 'SHIFT_IN_B' and 'CLKB (less than 5MHz in this system)' that the 64 channels per chip costed 12.8 μs at least. Therefore, the counting rates is less than 5 kHz and the corresponding maximum data transfer rate is less than 80 Mbps. The requirements for FEC of VA140 based SRS were showed in Table. 2.

### 3.3.2 ASIC cards and Adapter based on AGET

The second set of ASIC card and corresponding Adapter was designed based on AGET chip [15], which was specific developed for Generic Electronic system for Time Projection Chambers (GET) [21] [22]. The chip includes 64 channels and every channel consists of a charge sensitive preamplifier (CSA), a shaper, a discriminator and a 512-sample analog memory. The chip can be programmed to work in various mode with different charge ranges (120fC, 240fC, 1pC and 10pC), peaking time (16 values from 50 ns to 1 μs), sampling frequency (1 MHz to 100 MHz), and signal polarity (positive or negative). These characters gave it a wide variety roles in high energy physics experiments.

According to the AGET datasheet [15], the chip counting rates is less than 1 kHz. The chip needs 3.3V power supply, 3 pairs of differential controlling signals (TRIGGP, TRIGGM, WCKP, WCKM, RCKP, RCKM), and 6 single controlling signals (WRITE, READ, SC_DOUT, SC_DIN, SC_CK, SC_EN). The power consumption of AGET is about 10 mW/ch. To digitalize the outputs of AGET chips, a 14-bit quad-ADC (AD9259) was adopted to connect to 4 AGET chips, featuring up to 50 MSPS per channel with 7 pairs of differential controlling signals (ADC_CLKP, ADC_CLKN, DCO_P, DCO_N, FCO_P, FCO_N, and 4 pairs of ADC output signals), 4 single controlling signals (CSB, SDIO, SCLK, POWN) and 1 power supply (1.8V). With the help of some other auxiliary devices, the AGET ASIC card and Adapter were implemented.

In the AGET based SRS, the FEC with corresponding Adapter acted as a carrier for the "4-in-1" card. 4 AGET ASIC cards read out 256 channels of detector and connected to one Adapter. The required maximum data transfer rate which corresponds to the 1 kHz is less than 8 Mbps. The requirements for FEC of VA140 based SRS were showed in Table. 2.

### 3.3.3 The potential ASIC cards and Adapter based on APV25

The third set of ASIC card and corresponding Adapter can be designed based on APV25 [19], which is intended for read-out of silicon strip detectors in the CMS tracker. The chip contains 128 channels of preamplifier and shaper with a peaking time of 50ns and a charge range of 20 fC, driving a 192 column analogue memory into which samples are written at the 40MHz frequency.

According to the APV25 datasheet [19], the chip needs 1.25V and 2.5V power supplies, 2 pairs of differential controlling signals (TRIG+, TRIG-, CLK+, CLK-), and 4 single controlling signals (RST, SDAIN, SDOUT, SCLK). The power consumption of APV25 is about 2.3 mW/ch. To digitalize the outputs of APV25 chips, a 8-channel ADC (AD9637) is adopted to connect to 8 APV25 chips, featuring up to 40 MSPS per channel with 11 pairs of differential controlling signals (ADC_CLKP, ADC_CLKN, DCO_P, DCO_N, FCO_P, FCO_N, and 8 pairs of ADC output signals), 4 single controlling signals (CSB, SDIO, SCLK, POWN) and 1 power supply (1.8V). With the help of some other auxiliary devices, the APV25 ASIC card and Adapter can be implemented.

In the APV25 based SRS, the FEC with corresponding Adapter acted as a carrier for the "8-in-1" card. 8 APV25 ASIC cards read out 1024 channels of detector and connected to one Adapter. The requirements of APV25 based SRS for PCIe interface were showed in Table. 2.

## 4. Test Results

The prototype SRS with two sets of ASIC cards and Adapters based on VA140 and AGET were implemented. And they were assembled to work respectively. In this section, the performance of VA140 and AGET was tested firstly, then the transfer performance of FEC based on SiTCP and the power consumption.

Due to the VA140 and AGET chips both having calibration capacitors on chip, a general testing method was taken: a waveform generator (Tektronix, AFG3252) with attenuator was used to generate step pulses with different amplitudes. When the step pulses were applied to the on-chip capacitor, a certain amount of charge, which covered the full range, was injected into the selected channel of VA140 or AGET for performance testing.

In the VA140 based SRS, step pulses with amplitude ranging from 2 to 120 mV and 100 ns trailing edge were applied to the on-chip capacitor (2 pF), a certain amount of charge, from 4 fC to 240 fC which covered the full range, was injected into the 32 channels of VA140 respectively. Typical relationship between output peak value and input charge of the system was shown in Fig 7 (a). The quadratic curve indicates that the integral nonlinearity (INL) was better than 1.5%. And the noise (RMS) was about 0.15 fC without connecting to the detector.

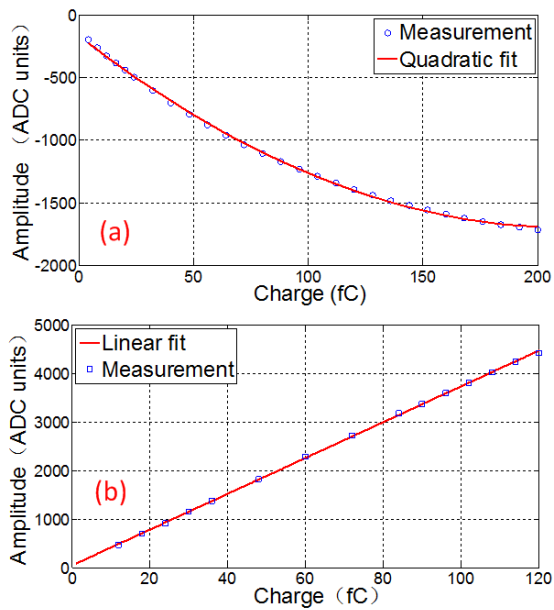

Fig. 7. (a) The quadratic fit result of VA140 with an INL of better than 1.5%. (b) The linear fit result of AGET with an INL of better than 2%.

In the AGET based SRS, the chip had a large number of operation modes. The charge measurement has been tested with the 120 fC range. The step pulses with amplitude ranging from 100 to 1000 mV and 100 ns trailing edge were applied to the on-chip capacitor (120 fF), a certain amount of charge, from 12 fC to 120 fC which covered the full range, was injected into the selected channel of AGET. The typical linearity curve of output peak value versus input charge was shown in Fig. 8 (b), with an INL of better than 2%. And the noise was better than 0.2 fC without connecting to the detector.

In order to test the transfer performance of FEC, the FEC was connected directly to a PC with a standard RJ-45 cable, acted as a TCP server and TCP client, respectively. The transfer speed can reach up to 530 Mbps per FEC. And the power consumption of the FEC was about 10 W (3.3V (2.8A), 5.0V (0.16A)).

## 5. Conclusion

A prototype SRS system based on VA140 and AGET have been designed and tested, which works well as designed. Took VA140 based SRS as an example, one chassis have an ability to assemble 17 FECs (another two slots for one power module), integrating 272 VA140 chips which handle 8704 detector channels. And if all of the 64 channels of

VA140 are bonded out in next work, one chassis can handle 17408 detector channels.

The FEC with a transfer speed up to 530 Mbps offers a maximum data transfer rate of about 9000 Mbps in one chassis and the flexible interface for other potential ASIC chips, expanding the system in application dimension.

*The authors thank Dr. Zhang Fei at the Institute of High Energy Physics, Chinese Academy of Sciences, for his useful suggestions and discussions.*